\documentclass[letterpaper, 10 pt, conference]
{IEEEtran}
\usepackage{amsmath}
\usepackage{amssymb}
\usepackage{amsfonts}
\usepackage{epsfig}
\usepackage{setspace}
\usepackage{cite}
\usepackage{calc}
\usepackage{color}
\usepackage[lmargin=0.75in,rmargin=0.75in,tmargin=1in,bmargin=0.85in]{geometry}

\title{Explicit Construction of Optimal Exact Regenerating Codes \\
for Distributed Storage}
\author{
{K. V. Rashmi{\small $^\dagger$}, Nihar B. Shah{\small $^\dagger$}, P. Vijay Kumar{\small $^\dagger$}, Kannan Ramchandran{\small $^\#$} }%
\vspace{1.6mm}\\
$^{^\dagger}$\,Dept. of ECE, Indian Institute Of Science, Bangalore, India. \\
Email: \{rashmikv, nihar, vijay\}@ece.iisc.ernet.in\\
$^{\#}$\,Dept. of EECS, University of California, Berkeley, USA. \\
Email: kannanr@eecs.berkeley.edu\\
}

\newtheorem{thm}{Theorem}

\newtheorem{lem}[thm]{Lemma}
\newtheorem{cor}[thm]{Corollary}

\newcommand{\beq}{\begin{equation}}
\newcommand{\eeq}{\end{equation}}

\newcommand{\bea}{\begin{eqnarray}}
\newcommand{\eea}{\end{eqnarray}}
\newcommand{\bean}{\begin{eqnarray*}}
\newcommand{\eean}{\end{eqnarray*}}
\newcommand{\bit}{\begin{itemize}}
\newcommand{\eit}{\end{itemize}}
\newcommand{\ben}{\begin{enumerate}}
\newcommand{\een}{\end{enumerate}}
\newcommand{\blem}{\begin{lem}}
\newcommand{\elem}{\end{lem}}
\newcommand{\bthm}{\begin{thm}}
\newcommand{\ethm}{\end{thm}}
\newcommand{\bpf}{\begin{proof}}
\newcommand{\epf}{\end{proof}}

\begin{document}
\maketitle\thispagestyle{empty}
\bibliographystyle{ieeetran}

\begin{abstract}

Erasure coding techniques are used  to increase the reliability of distributed storage systems while minimizing storage overhead. Also of interest is minimization of the bandwidth required to repair the system following a node failure. In a recent paper, Wu et al. characterize the tradeoff between the repair bandwidth and the amount of data stored per node. They also prove the existence of regenerating codes that achieve this tradeoff.

In this paper, we introduce {\em Exact Regenerating Codes}, which are regenerating codes possessing the additional property of being able to duplicate the data stored at a failed node. Such codes require low processing and communication overheads, making the system practical and easy to maintain. Explicit construction of exact regenerating codes is provided for the minimum bandwidth point on the storage-repair bandwidth tradeoff, relevant to distributed-mail-server applications. A subspace based approach is provided and shown to yield necessary and sufficient conditions on a linear code to possess the exact regeneration property as well as prove the uniqueness of our construction.

Also included in the paper, is an explicit construction of regenerating codes for the minimum storage point for parameters relevant to storage in peer-to-peer systems. This construction supports a variable number of nodes and can handle multiple, simultaneous node failures. All constructions given in the paper are of low complexity, requiring low field size in particular.

\end{abstract}

\section{Introduction\label{sec:intro}}

 Reliability is a major concern in large distributed
storage systems where data is stored across multiple unreliable
storage nodes. It is well known that adding redundancy increases the
reliability of the system but at the cost of increased storage.
Erasure coding based techniques \cite{oceanstore},\cite{totalRecall}
(eg. using Maximum distance separable(MDS) codes) have been used to
minimize this storage overhead.

In a distributed storage system, when a subset of the nodes fail,
the system needs to repair itself using the existing nodes. In
erasure coding based systems, each node stores a fragment of an MDS
code. Upon failure of a node, the failed fragment can be restored
back using the existing fragments. The amount of data that needs to
be downloaded to restore the system after a node failure is one of
the significant parameters of a distributed storage system. In
\cite{DimKan1} the authors introduce a new scheme called
Regenerating Codes which store a larger amount of data at each node compared to an MDS code,
in order to reduce the repair bandwidth. In \cite{YunDimKan} the
authors establish a tradeoff between the amount of storage required
at each node and the repair bandwidth. Two extreme and practically
relevant points on this storage-repair bandwidth tradeoff curve are
the minimum bandwidth regeneration(MBR) point which represents the
operating point with least possible repair bandwidth, and the
minimum storage regeneration(MSR) point which corresponds to the
least possible amount of data stored at a node. By an optimal
Regenerating Code, we will mean a Regenerating Code that meets the
storage-repair bandwidth tradeoff.

A principal concern in the practical implementation of distributed
storage codes is computational complexity. A practical study of the
same has been carried out in \cite{Complexi} for random linear
regenerating codes. Although the existence of optimal regenerating
codes was proved in \cite{YunDimKan}, for code construction, the
authors have suggested the use of a general network-coding-based
code construction algorithm due to Jaggi et al \cite{Jaggi}. The
drawbacks of such an approach include high complexity of code
construction as well as the requirement of a large field size.

In this paper, we introduce \emph{Exact Regenerating Codes}, which
are regenerating codes possessing the additional property of being
able to regenerate back the same node upon failure. We give a
low-field-size, low-complexity, explicit construction for exact
regenerating codes at the MBR point. Using the subspace based approach
provided, we also prove that our code is unique among all the linear
codes for this point. Explicit construction is also given for
regenerating codes at the MSR point for suitable parameters which
can handle multiple node failures. To the best of our knowledge, our
codes are the first explicit constructions of optimal regenerating
codes.

In \cite{YunDim}, Wu et al. also independently introduce the notion
of exact regeneration(termed exact repair in \cite{YunDim}) for the
MSR point. However, the codes introduced in their work do not meet
the storage-repair bandwidth tradeoff. The construction proposed by
them is of high complexity and also has the disadvantage of a large
field size requirement.

The rest of the paper is organized as follows. In Section \ref{sec:renew} we introduce the notion of Exact Regenerating Codes. Explicit construction of Exact Regenerating Codes for the MBR point is given in Section \ref{sec:MBR}. The complexity and the field size requirement of the proposed code construction algorithm are also analyzed here. In Section \ref{sec:subspaceview}, a subspace based approach to construction of these codes is provided which is later used to prove the uniqueness of our construction. Section \ref{sec:MSR} provides a construction of regenerating codes for the MSR point for certain practically relevant parameters. Finally,
conclusions are drawn in Section \ref{sec:conclude}.

\section{Exact Regenerating Codes} \label{sec:renew}

\begin{figure}[h]
\vspace{-10pt}
  \centering
    \includegraphics[width=0.4\textwidth]{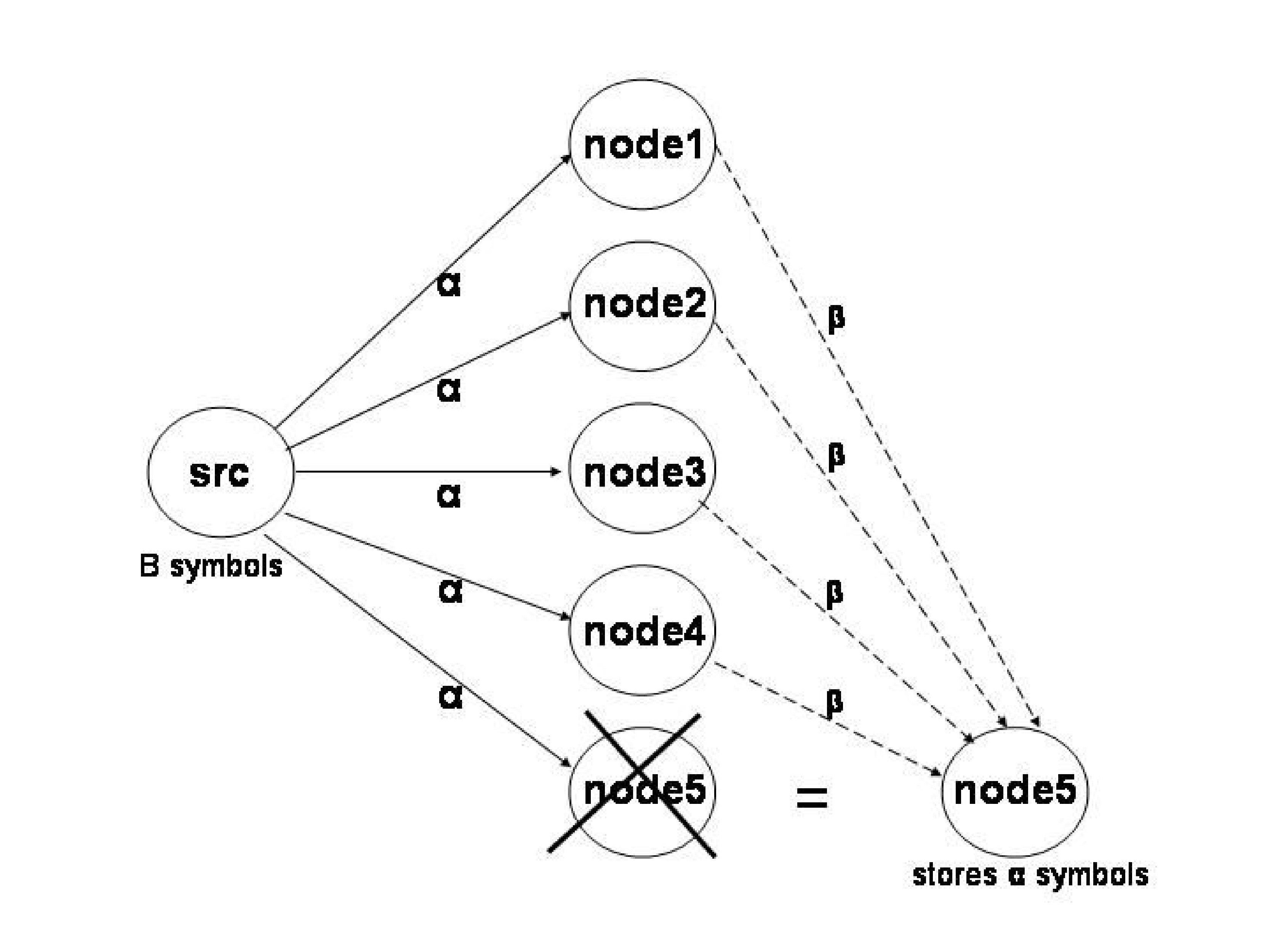}
    \vspace{-15pt}
\caption{An illustration of exact regeneration: On failure of node
5, data from nodes 1 to 4 is used to regenerate back the same data
that node 5 earlier had.} \label{fig:EgExact}
    \end{figure}

The system description is as follows. All data elements belong to a
finite field $\mathbb{F}_q$ of size $q$. The total size of the file
is $B$ units. The data is stored across $n$ storage nodes in a
distributed manner where each node can store up to $\alpha$ units of
data. A data collector(DC) connects to any $k$ out of these $n$
nodes to reconstruct the entire data. This property is termed as
`reconstruction property'. The data collector is assumed to have
infinite capacity links so that it downloads all the data stored in
these $k$ nodes.

When a node fails, a new node is added in its place by downloading $\beta$ units of data from any $d (\geq k)$\footnote{From \cite{YunDimKan}, if $d < k$, the mincut condition will require data reconstruction to hold for $d$ nodes, hence $k$ can be set as $d$.} out of the remaining $n-1$ nodes. In regenerating codes as introduced in \cite{YunDimKan}, the new node formed need not be identical to the failed one. It should satisfy the reconstruction property along with the existing nodes. This property wherein a new node satisfying reconstruction can be created as a replacement for a failed node is termed as `regeneration property'. Any other node subsequently regenerated using this node should satisfy both the properties. Hence the new node along with all other nodes should satisfy these properties for a possibly infinite sequence of failures and regenerations.

We introduce a desirable property into regenerating codes wherein
the regenerated node is identical to the one which failed. We
will call regenerating codes having this additional property as
`Exact Regenerating Codes'. Fig. \ref{fig:EgExact} shows an example
of the this scheme. As a failed node is replaced by an identical
node, Exact Regenerating Codes have to satisfy the reconstruction
property at only one level. Also, the additional communication and
processing overheads required to update all the other nodes and data
collectors about the new node is completely avoided. This makes the
storage system practical and easy to maintain.

\section{Exact Regenerating codes for the MBR point}\label{sec:MBR}

The MBR point is the fastest recovery point (on the storage-repair
bandwidth tradeoff curve) in terms of the data to be downloaded for
regeneration per unit of the source data. Also, among all the
possible values of $d$, $d=n-1$ point gives the fastest recovery as
all the existing nodes simultaneously help in the regeneration of
the failed node. Hence the MBR point with $d=n-1$ is highly suitable
for applications such as distributed mail servers, where it is
crucial to restore the system in the shortest possible time.

This section gives the construction of linear exact regenerating
codes at the MBR point for $d=n-1$ and any $k$. At the MBR point, optimal $\alpha$ and $\beta$ on the storage-repair
bandwidth tradeoff curve are given by (from \cite{YunDimKan}):

\bea
\label{eqn:MBR}
 (\alpha_{MBR},\beta_{MBR})= \left( \frac{2Bd}{2kd-k^2+k}, \frac{2B}{2kd-k^2+k}
 \right)
\eea

Clearly for a feasible system we need $\beta$ to be an
integer\footnote{It can be seen from equation (\ref{eqn:MBR}) that
if $\beta$ is an integer, then $\alpha$ and $B$ are also integers.}.
Assume $\beta$ to be the smallest possible positive integer, i.e.
$\beta=1$. Then we have \bea B = kd - \frac{k(k-1)}{2}\eea and \bea
\alpha = d \eea

For any larger file size, the source file is split into chunks of
size $B$, each of which can be separately solved using the
construction for $\beta=1$. Reconstruction and regeneration will
be performed separately on these smaller chunks and hence additional
processing and storage required to perform these operations is greatly reduced.

\subsection{Code construction}
Denote the source symbols of the file by $\underline{f} = (f_{0} \;
f_{1} \; f_{2} \; \ldots \; f_{B-1})^t$. Let $d=n-1$ and
$\theta=\frac{d(d+1)}{2}$. Let V be a $n$ x $\theta$ matrix with the
following properties:
\begin{enumerate}
\item Each element is either $0$ or $1$.
\item Each row has exactly $d$ $1$'s.
\item Each column has exactly two $1$'s.
\item Any two rows have exactly one intersection of $1$'s.
\end{enumerate}

It is easy to see that V is the incidence matrix of a fully
connected undirected graph with $n$ vertices. Our construction of
exact regenerating codes for the MBR point uses the above described
matrix V. Consider a set of $\theta$ vectors $\{\underline{v}_1, \;
\underline{v}_2,\ldots,\underline{v}_\theta\}$ which form a
$B$-dimensional MDS code. The vectors
$\underline{v}_i\;(i=1,\ldots,\theta)$ are of length $B$ with the
constituent elements taken from the field $\mathbb{F}_q$. Node $j$
stores the symbol $\underline{f}^t \underline{v}_i$ if and only if
V$(j,i) = 1$. Thus in the graph corresponding to V, vertices represent the nodes, and edges represent the vectors corresponding to the symbols stored.  Thus, by the properties of the matrix V, we get $n$
nodes each storing $d(=\alpha)$ symbols. Properties 3 and 4 ensure
that each row intersects every other row in distinct columns. The
validity of this code as a exact regenerating code for the MBR point
is shown below.
\newline
\begin{figure}[t]
\vspace{-10pt}
  \centering
	\includegraphics[width=2in]{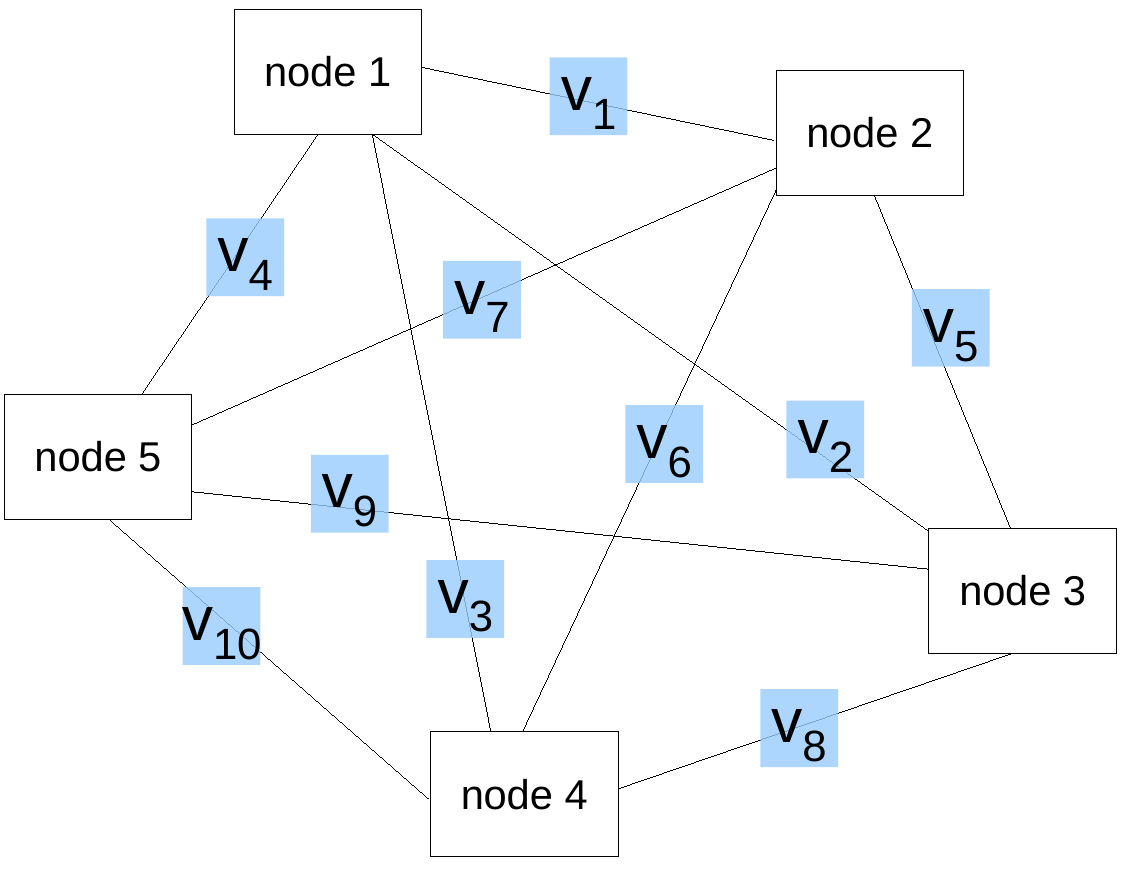}
    \vspace{-10pt}
\caption{Fully connected undirected graph with $5$ vertices. Vertices represent nodes and edges represent vectors corresponding to the common symbol between two nodes.} \label{fig:complete}
    \end{figure}

\textit{Data Reconstruction:} The DC connects to any $k$ out of the
$n$ storage nodes and downloads all the $k\alpha$ symbols stored. As
any two rows of the matrix V intersect in only one column and any row intersects all other rows in distinct columns, out of the $k\alpha$ symbols downloaded, exactly ${k\choose2}$ symbols are repetitions and do not
add any value. Hence the DC has $k\alpha - {k\choose2} = B$ distinct
symbols of a $B$-dimensional MDS code, using which the values of the
source symbols $f_{0},\ldots,f_{B-1}$ can be easily obtained.
\newline

\textit{Exact Regeneration:} The matrix V provides a special
structure to the code which helps in exact regeneration. Properties
3 and 4 of the matrix V imply that the each of the existing $n-1$ nodes contain
one distinct symbol of the failed node. Thus exact regeneration of the
failed node is possible by downloading one symbol each from the
remaining $n-1$ nodes.
\newline

In section \ref{sec:subspaceview} it will be proved that this code
construction scheme is unique for linear exact regenerating codes up
to the choice of vectors $\{\underline{v}_1, \; \underline{v}_2, \;
\ldots \; , \; \underline{v}_\theta\}$. In the above description we
have chosen these set of vectors to form a $B$-dimensional MDS code.
In fact, it suffices if the vectors are chosen such that, for any
set of $k$ nodes, the constituent vectors are linearly independent.

\subsection{Example}

Let $n=5,\; k=3$. We get $d=n-1=4$ and $\theta = 10$. Putting $\beta = 1$ gives
$\alpha = 4$ and $B = 9$. As described in the
previous section, the matrix V is the incidence matrix of a fully
connected undirected graph with $5$ vertices (Fig. \ref{fig:complete}) as given below:

\vspace{5pt}
 \noindent
\begin{tabular}{|c|*{10}{c}|}
\hline
   &v1&v2&v3 & v4 & v5 & v6 & v7 & v8 & v9 & v10 \\
\hline
   n1&1&1 & 1 & 1 & 0 & 0 & 0 & 0 & 0 & 0\\
   n2&1&0 & 0 & 0 & 1 & 1 & 1 & 0 & 0 & 0\\
   n3&0&1 & 0 & 0 & 1 & 0 & 0 & 1 & 1 & 0\\
   n4&0&0 & 1 & 0 & 0 & 1 & 0 & 1 & 0 & 1\\
   n5&0&0 & 0 & 1 & 0 & 0 & 1 & 0 & 1 & 1\\
  \hline
\end{tabular}
\vspace{5pt}
\newline
\noindent Thus the $5$ nodes store the following symbols:
\newline
Node 1: $\{\underline{f}^t \underline{v}_1, \;  \underline{f}^t
\underline{v}_2, \;  \underline{f}^t \underline{v}_3, \;
\underline{f}^t \underline{v}_4\}$\newline Node 2:
$\{\underline{f}^t \underline{v}_1, \;  \underline{f}^t
\underline{v}_5, \;  \underline{f}^t \underline{v}_6, \;
\underline{f}^t \underline{v}_7\}$\newline Node 3:
$\{\underline{f}^t \underline{v}_2, \;  \underline{f}^t
\underline{v}_5, \;  \underline{f}^t \underline{v}_8, \;
\underline{f}^t \underline{v}_9\}$\newline Node 4:
$\{\underline{f}^t \underline{v}_3, \;  \underline{f}^t
\underline{v}_6, \;  \underline{f}^t \underline{v}_8, \;
\underline{f}^t \underline{v}_{10}\}$\newline Node 5:
$\{\underline{f}^t \underline{v}_4, \;  \underline{f}^t
\underline{v}_7, \;  \underline{f}^t \underline{v}_9, \;
\underline{f}^t \underline{v}_{10}\}$
\newline

\textit{Reconstruction: } Suppose the data collector connects to
nodes $1$, $2$ and $3$. It can retrieve the symbols $\underline{f}^t
\underline{v}_1$,\ldots,$\underline{f}^t \underline{v}_9$, and using
these, recover the source symbols $f_{0},\ldots,f_{8}$. The same
holds for any choice of $3$ nodes.

\textit{Regeneration:} Suppose node $3$ fails. Then, node $1$ gives
$\underline{f}^t \underline{v}_2$, node $2$ gives $\underline{f}^t
\underline{v}_5$, node $4$ gives $\underline{f}^t \underline{v}_8$
and node $5$ gives $\underline{f}^t \underline{v}_9$. All these four
symbols are stored as the new node $3$. Thus the regenerated node $3$ stores exactly the same symbols as the failed node.

In this example, $\theta = B + 1$ and hence we can take the vectors
$\{\underline{v}_{1},\ldots,\underline{v}_{10}\}$ to form a single
parity check code of dimension $9$. So the exact regenerating code
for this set of parameters can be obtained in $\mathbb{F}_2$.

\subsection{Field size required}
The required field size is the minimum field size required to
construct a [$\theta$,\;$B$] MDS code. If we use a Reed-Solomon
code, the minimum field size required for our construction turns out to
be $\theta(=n(n-1)/2)$. In \cite{YunDimKan} authors have suggested
to cast the problem of constructing deterministic regenerating codes
as a virtual multicast network code construction problem and then
use the algorithm due to Jaggi et al. \cite{Jaggi} to determine the
network coefficients. This algorithm requires field size of the
order of number of sinks, which in this case leads to a very high
field size. In fact, the problem of exact regenerating code
construction leads to a non-multicast network code problem for which
there are very few results available
\cite{lehman},\cite{multUnicast}.

\subsection{Complexity}

\textit{Code construction: } Code construction is immediate given the incidence matrix V of a fully connected graph with $n$ vertices. No arithmetic operations are required.

\textit{Node Regeneration: } The method used for regeneration does
not require the existing nodes to perform any additional operations.
Each existing node just has to pass one symbol to the new node from the set of $\alpha$ symbols stored in it.

If the regeneration is not exact, additional communication to the
nodes and data collectors about changes in the code coefficients is
necessary. Also, all the nodes need to recalculate the vectors which they have to
pass for subsequent regenerations. In the case of exact
regeneration, these overheads are avoided.

\textit{Data Reconstruction: } To facilitate the DC to easily decode
the downloaded data, one set of $k$ nodes can be made systematic,
i.e. these k nodes will store the source symbols without any encoding. This can be achieved by performing a change of
basis on the $B$-dimensional vector space spanned by the vectors
$\{\underline{v}_1, \underline{v}_2,\ldots,\underline{v}_\theta\}$,
so that the desired $k$ nodes have entire data in uncoded form.
Hence, if the DC preferably connects to this set of $k$ nodes, no
decoding is necessary.

If regeneration is not exact, the systematic property cannot be
maintained. When any of one of the $k$ nodes chosen to be systematic
fails, the regenerated node may not be in the systematic form and
hence the property will be lost.

\section{Subspace viewpoint and uniqueness}\label{sec:subspaceview}

In the construction of exact regenerating codes given in section
\ref{sec:MBR}, nodes were viewed to be storing $\alpha$ symbols each
from a finite field. In this section, we provide an alternative
viewpoint based on subspaces which completely characterizes linear
exact regenerating codes for the MBR point for any values of
$(n,\;k,\;d)$. By a linear storage code, we mean that any symbol
stored is a linear combination of the source symbols, and only
linear operations are allowed on them.

The subspace viewpoint will be used to prove the necessary and
sufficient conditions for a linear storage code to be an exact
regenerating code for the MBR point. This subsequently leads to the uniqueness of our
construction.

Define a vector $\underline{f}$ of length $B$ consisting of the
source symbols (as in section \ref{sec:MBR}). Since each source
symbol can independently take values from $\mathbb{F}_q$, the $B$
source symbols can be thought of as forming a $B$-dimensional vector
space over $\mathbb{F}_q$.

Since the code is linear, any stored symbol can be written as
$\underline{f}^t\underline{\ell}$ for some vector
$\underline{\ell}$. These vectors which specify the linear
combinations define the code, and the actual symbols stored depend
on the instantiation of $\underline{f}$. Since a node stores
$\alpha$ symbols, it can be considered as storing $\alpha$ vectors
of the code, i.e. node $i$ stores the vectors
$\underline{\ell}^{(i)}_1,\ldots,\underline{\ell}^{(i)}_\alpha$.
Linear operations performed on the stored symbols are equivalent to
the same operations performed on these vectors. Hence we say that
each node \textit{stores a subspace} of dimension at most $\alpha$
i.e. \beq \text{node }i \text{: } W_i
=\left<\underline{\ell}^{(i)}_1,\ldots,\underline{\ell}^{(i)}_\alpha\right>
\nonumber \eeq where $W_i$ denotes the subspace stored in node $i$ ,
$i=1,\ldots,n$ and $\left<.\right>$ indicates the span of vectors.

For regeneration of a failed node, $d$ other nodes provide $\beta$
symbols each. We say that each node \textit{passes a subspace} of
dimension at most $\beta$.

Consider the exact regeneration of some node $i$ using any $d$ out
of the remaining $n-1$ nodes. Denote this set of $d$ nodes by
$\mathbf{D}$, and let $j\; \in\;\mathbf{D}$. Let
$S^{(i)}_{j,\mathbf{D}}$ denote the subspace passed by node $j$ for
the regeneration of node $i$.

In the following lemmas, we prove certain subsapce properties associated with linear exact regenerating codes at the MBR point.

\begin{lem}\label{lem:alphaDim}
For any $(n,\;k,\;d)$ linear exact regenerating code for the MBR
point, each node stores an $\alpha$-dimensional subspace, i.e. \beq
dim\{W_i\} = \alpha, \;\;\; \forall i\; \in
\;\{1,\ldots,n\}.\nonumber \eeq
\end{lem}

\begin{proof}
Consider data reconstruction by a DC connecting to any $k$ nodes,
$\Lambda_1$,\ldots,$\Lambda_k$. Let these $k$ nodes store subspaces
with dimensions $\Omega_1$,\ldots,$\Omega_k$ respectively. As each
node can store a subspace of dimesion at most $\alpha$, \beq
\Omega_i \leq \alpha,\;\;\forall i\;\in\; \{1,\ldots,k\}
\label{eqn:dimspace} \eeq For the DC to be able to reconstruct all
the data, the dimension of the sum space of these $k$ subspaces
should be $B$, i.e. \beq \label{eqn:unionDim} dim\{W_{\Lambda_1} +
W_{\Lambda_2} + \cdots + W_{\Lambda_k}\} = B \eeq
\newline
Using the expression for the dimension of sum of two subspaces
recursively we get, \bea
dim&&\hspace{-20pt}\{W_{\Lambda_1}+\cdots+W_{\Lambda_k}\} \nonumber \\
&&\hspace{-35pt} =dim \{W_{\Lambda_1}\}+ dim \{W_{\Lambda_2}\}-dim \{W_{\Lambda_1} \cap W_{\Lambda_2}\} \nonumber \\
    &&\hspace{-30pt}+ dim \{W_{\Lambda_3}\} - dim \{W_{\Lambda_3} \cap \{W_{\Lambda_1} + W_{\Lambda_2}\}\} \nonumber \\
    &&\hspace{-30pt}\cdots \nonumber \\
    &&\hspace{-30pt}+dim\{W_{\Lambda_k}\}-dim\{W_{\Lambda_k} \cap \{W_{\Lambda_1} +\cdots+ W_{\Lambda_{k-1}}\} \} \nonumber \hspace{-20pt}\\
&&\hspace{-35pt}=\sum_{i=1}^{k}dim\{W_{\Lambda_i}\} \nonumber \\
&&\hspace{-30pt}-dim\{W_{\Lambda_k} \cap \{W_{\Lambda_{k-1}}+\cdots+W_{\Lambda_1}\} \} \nonumber \\
&&\hspace{-30pt}-\cdots-dim \{W_{\Lambda_3} \cap \{ W_{\Lambda_2}+W_{\Lambda_1}\}\} \nonumber \\
&&\hspace{-30pt}-dim\{W_{\Lambda_2} \cap W_{\Lambda_1}\} \label{eqn:dup1} \\
\label{eqn:dup2} &&\hspace{-35pt}\leq  \sum_{i=1}^{k}\Omega_i \nonumber \\
&&\hspace{-30pt}-(\Omega_k - (d-(k-1))\beta)^+ \nonumber \\
&&\hspace{-30pt}-\cdots-(\Omega_3-(d-2)\beta)^+ \nonumber \\
&&\hspace{-30pt}-(\Omega_2 - (d-1)\beta)^+    
\\\label{eqn:dup3} &&\hspace{-35pt}\leq\Omega_1 +\sum_{l=2}^k(d - (l-1))\beta \\&&\hspace{-35pt}= \Omega_1 +  (k-1)d\beta - \{k-1+\cdots+2+\;1\}\beta \\
&&\hspace{-35pt}= \Omega_1 - \alpha +B \label{eqn:dup4}\\&&\hspace{-35pt}\leq B\label{eqn:dup5}
\eea

In (\ref{eqn:dup2}), $(x)^+$ stands for $max(x,0)$. The
justification for (\ref{eqn:dup2}) is as follows. Suppose nodes
$\Lambda_1,\ldots,\Lambda_{l-1}$ and some other $(d-(l-1))$ nodes
participate in the regeneration of node $\Lambda_l$. The maximum
number of linearly independent vectors that the $(d-(l-1))$ nodes
(other than $\Lambda_1,\ldots,\Lambda_{l-1}$) can contribute is
$(d-(l-1))\beta$. If this quantity is less than $\Omega_l$ then the
$l-1$ nodes under consideration will have to pass the remaining
dimensions
to node $l$. Hence for any $l=2,\ldots,k$ \bea \hspace{-25pt}dim\{W_{\Lambda_l}\cap\{W_{\Lambda_{l-1}}+\cdots+W_{\Lambda_1}\}\}\nonumber&&\\
&\hspace{-80pt}\geq& \hspace{-40pt}(\Omega_l - (d-(l-1))\beta)^+
\label{eq:minVal} \eea

Equation (\ref{eqn:dup3}) follows by the property that any two non-
negative numbers $y_1$ and $y_2$ satisfy the inequality $(y_1 - (y_1
- y_2)^+) \leq y_2$. Equation (\ref{eqn:dup4}) follows from
(\ref{eqn:MBR}) and equation (\ref{eqn:dup5}) from
(\ref{eqn:dimspace}). Now, for equation (\ref{eqn:unionDim}) to
hold, (\ref{eqn:dup5}) should be satisfied with equality, which
forces $\Omega_1 = \alpha$. Similarly, expanding with respect to
other nodes, and considering different sets of $k$ nodes, we get
$dim\{W_i\} = \alpha$, $\forall i\;\in\;\{1,\ldots,n$\}.
\end{proof}

\begin{cor}
\label{cor:multIntersect} Let $\mathbf{D}_m$ be any subset of
$\mathbf{D}$ of size $m$, where $m < k$. For any $(n,\;k,\;d)$
linear exact regenerating code at the MBR point, \beq  dim
\left\{W_i \cap \left\{\sum_{j \in \{\mathbf{D}_m\}}
W_j\right\}\right\}  = m\beta\nonumber\eeq.
\end{cor}

\begin{proof}
Putting $\Omega_l = \alpha = d\beta$ in (\ref{eq:minVal}) we get,
\bea \hspace{+10pt}dim(W_{\Lambda_l} \cap
\{W_{\Lambda_{l-1}}+\cdots+W_{\Lambda_1} \})\;\geq\;(l-1)\beta
\label{eq:InterMinVal}\eea Using (\ref{eqn:dup1}) and
(\ref{eq:InterMinVal}), \bea
dim&&\hspace{-25pt}\{W_{\Lambda_1}+\cdots+W_{\Lambda_k}\} \nonumber\\
&=&\sum_{i=1}^{k}dim\{W_{\Lambda_i}\} \nonumber \\
&&-dim\{W_{\Lambda_k} \cap \{W_{\Lambda_{k-1}}+\cdots+W_{\Lambda_1}\} \} \nonumber \\
&&-\cdots-dim \{W_{\Lambda_3} \cap \{ W_{\Lambda_2}+W_{\Lambda_1}\}\} \nonumber \\
&&-dim\{W_{\Lambda_2} \cap W_{\Lambda_1}\} \\
\label{eqn:cordup}
&\leq& k\alpha - \{k-1+\cdots+2+\;1\}\beta\label{eqn:cordup1} \\
\label{eqn:cordup3} &=& B \eea For equation (\ref{eqn:unionDim}) to
hold, (\ref{eqn:cordup}) should be satisfied with equality. This
along with (\ref{eq:InterMinVal}) gives the result.
\end{proof}

Note that putting $m=1$ gives \beq dim\{W_i \cap W_j\} = \beta
\label{eq:intersectBeta}\eeq

\begin{lem}
\label{lem:oneIntersect} For any $(n,\;k,\;d)$ linear exact
regenerating code at the MBR point, \beq  S^{(i)}_{j,\mathbf{D}} =
W_i \cap W_j\nonumber \eeq Also, the subspaces $W_i \cap W_j$ are
linearly independent $\forall\; j\; \in \; \mathbf{D}$.
\end{lem}

\begin{proof}
Consider the exact regeneration property of node $i$. As $d\beta =
\alpha$, node $i$ must store all the information passed by the nodes
in $\mathbf{D}$. Hence, the subspace passed by node $j$ must be a
subspace of $W_i$ as well, i.e. \beq \label{eq:oneIntersect_subset}
S^{(i)}_{j,\mathbf{D}} \subseteq (W_i \cap W_j) \eeq

Also, \bea \sum_{j \in \{\mathbf{D}\}} dim\left\{
S^{(i)}_{j,D}\right\}&\geq&dim\left\{\sum_{j \in \{\mathbf{D}\}}
S^{(i)}_{j,D}\right\} \label{eq:oneIntersect_dimsumspace}\\
&=&dim\left\{W_i\right\}\\&=&d\beta \eea which along with the fact
that $dim\{S^{(i)}_{j,\mathbf{D}}\} \leq \beta$ implies that
equation (\ref{eq:oneIntersect_dimsumspace}) should be satisfied
with equality and \beq
 dim\{S^{(i)}_{j,D}\} = \beta \label{eq:intersectBetaS}\eeq
From equations (\ref{eq:intersectBeta}),
(\ref{eq:oneIntersect_subset}) and (\ref{eq:intersectBetaS}), it
follows that \beq S^{(i)}_{j,\mathbf{D}} = W_i \cap W_j \eeq
Equality of equation (\ref{eq:oneIntersect_dimsumspace}) implies
that the subspaces $S^{(i)}_{j,\mathbf{D}}$ are linearly independent
$\forall\; j\; \in \; \mathbf{D}$.\end{proof}

Hence for any linear exact regenerating code for the MBR point, each
node should store an $\alpha$ dimensional subspace, and the
intersection subspaces of a node with any $d$ other nodes should
have dimension $\beta$ each and should be linearly independent.

The following theorems prove the uniqueness of our code for the MBR
point.

\begin{thm}
\label{thm:unique} Any linear exact regenerating code for the MBR
point with $d = n - 1$ should have the same subspace properties as our code and hence the same structure as our code.
\end{thm}

\begin{proof}
Let $C$ be an exact regenerating code obtained via our construction.
Let $C^\prime$ be another optimal exact regenerating code for the MBR point which satisfies the
reconstruction and exact regeneration properties. Let $W_1^\prime,
\ldots, W_n^\prime$ be the subspaces stored in nodes $1, \ldots,n$
respectively in code $C^\prime$. Apply Lemma \ref{lem:oneIntersect}
to node $1$ in $C^\prime$ and let $s_2^\prime,\ldots,s_n^\prime$ be
the $\beta$-dimensional intersection subspaces of node $1$ with
nodes $2, \ldots ,n$ respectively. As $s_2^\prime,
\ldots,s_n^\prime$ are linearly independent subspaces spanning
$\alpha$ dimensions, they constitute a basis for $W_1^\prime$ and
hence can be replaced as the contents of node $1$. Now consider node
$2$. One of the intersection subspaces will be $s_2^\prime$ (with
node $1$). Let $s_3^{\prime\prime}, \ldots,s_n^{\prime\prime}$ be
the intersection subspaces of node 2 with nodes $3,\ldots,n$. Again,
$s_2^\prime$ and $s_3^{\prime\prime}, \ldots, s_n^{\prime\prime}$
form a basis for $W_2^\prime$ and hence node $2$ can be replaced by
these. Continuing in the same manner across all the remaining nodes,
it is easy to see that the code $C^\prime$ has the same structure as $C$.
\end{proof}

Hence our code is unique upto the choice of basis for the node subspaces.

\begin{thm}
\label{cor:otherparams} A necessary and sufficient condition for any
linear code to be $(n,k,d)$ exact regenerating code for the MBR
point is that any set of $d+1$ nodes should have the same structure as
our code.
\end{thm}

\begin{proof}\textit{Necessity: } If there exists a linear exact regenerating
code at the MBR point for some $(n,k,d)$, then any set of $d+1$
nodes from this code should work as a code for the parameters
$(d+1,k,d)$. Hence, from Theorem \ref{thm:unique}, any set of $d+1$
nodes is of the same structure as our code.

\textit{Sufficiency: } Suppose there exists a linear code such that
any set of $d+1$ nodes from this code has the same structure as our code.
Consider a DC connecting to some $k$ nodes. This set of $k$ nodes
can be viewed as a subset of some $d+1$ nodes which will have the
same structure as our code. Hence, the DC can reconstruct the entire
data. Consider a failed node, and some $d$ nodes used to regenerate
it. Since this set of $d+1$ nodes will also have the same structure as
our code, exact regeneration of the failed node is possible. Thus,
reconstruction and exact regeneration properties are established.
\end{proof}

\section{Regenerating Codes for the MSR point}\label{sec:MSR}

The MSR point requires the least possible storage at the nodes (with respect to
the storage-repair bandwidth tradeoff curve). This operating point
particularly suits applications like storage in peer-to-peer systems
where storage capacity available from the participating nodes is
very low. In such systems, multiple node failures are quite frequent
as nodes enter and exit the system at their own will. Hence the
system should be capable of regenerating a failed node using only a
small number of existing nodes. Also, the number of nodes in the
system changes dynamically. Hence the code should work even if the
number of nodes keeps varying with time.

In this section we give an explicit construction for regenerating
codes at the MSR point for $d = k+1$ and any $n$. This set of
parameters makes the code capable of handling any number of failures
provided that at least $k+1$ nodes remain functional. Note that, by
definition, if less than $k$ nodes are functional then a part of the
data will be permanently lost. If exactly $k$ nodes are functional,
then these nodes will have to pass all the information stored in
them for regeneration, hence no optimization of the repair bandwidth
is possible.

 At the minimum storage point, optimal
$\alpha$ and $\beta$ on the storage-repair bandwidth tradeoff curve
are given by (from \cite{YunDimKan}):

\bea
 (\alpha_{MSR},\beta_{MSR})= \left( \frac{B}{k}, \frac{B}{k(d-k+1)}
 \right)
 \label{eq:alpha}
\eea

By the same argument as in the MBR case, we choose $\beta=1$ for our
construction, which gives \beq B=k(d-k+1) \label{eq:B} \eeq and \beq
\alpha=d-k+1 \label{eq:alphaMSR}\eeq

\subsection{Code construction:}
With $d=k+1$, from equations (\ref{eq:B}) and (\ref{eq:alphaMSR}) we
have \beq B=2k \eeq  and \beq \alpha=2 \eeq Partition the source
symbols into two sets: $f_0,\ldots,f_{k-1},$ and
$g_0,\ldots,g_{k-1}$. Let $\underline{f}^t=(f_{0} \; f_{1} \; \ldots
\; f_{k-1}), \text{ and } \underline{g}^t=(g_0 \; g_1 \; \ldots\;
g_{k-1} )$.

Node $i\;( i=1,\ldots,n)$ stores $(\underline{f}^t\underline{p}_i \;
, \; \underline{g}^t\underline{p}_i+\underline{f}^t\underline{u}_i)$
as its two symbols. We shall refer to the vectors $\underline{p}_i$
and $\underline{u}_i$ as the \textit{main vector} and the
\textit{auxiliary vector} of a node respectively. The elements of
the auxiliary vectors are known but can take any arbitrary values
from $\mathbb{F}_q$. The main vectors are the ones which are
actually used for reconstruction and regeneration.

Let the set of main vectors $\underline{p}_i(i=1,\ldots,n)$ form a
$k$-dimensional MDS code over $\mathbb{F}_q$. The required field size is the minimum field size required to construct an $[n,k]$
MDS code. If we use a Reed-Solomon code, the minimum field size
required turns out to be just $n$.

For example, consider $n=5, \;k=3 \text{ and } d=4 $. We have $B=6$
and $f_{0},\;f_{1},\;f_{2},\;g_{0},\;g_{1} \text{ and }g_{2}$ as the
source symbols. Let the main vectors $\underline{p}_i\;(
i=1,\ldots,n)$ form a Reed-Solomon code, with $\underline{p}_i = (1
\; \theta_i \; \theta^{2}_i)^t$. $\theta_i \;(i=1,\ldots,5)$ take
distinct values from $\mathbb{F}_q (q \geq 5)$. We can initialize
elements of $\underline{u}_i (i=1,\ldots,5)$ to any arbitrary values
from $\mathbb{F}_q$.

\subsection{Reconstruction:}
A data collector will connect to any $k$ nodes and download both the
symbols stored in each of these nodes. The first symbols of the $k$
nodes provide $\underline{f}^t\underline{p}_i$ at $k$ different
values of $i$. To solve for $\underline{f}$, we have $k$ linear
equations in $k$ unknowns. Since $\underline{p}_i$'s form a
$k-$dimensional MDS code, these equations are linearly independent,
and can be solved easily to obtain the values of $f_0,\ldots,f_{k-1}$ .

Now, as $\underline{f} \; \text{and} \; \underline{u}_i$ are known,
$\underline{f}^t\underline{u}_i$ can be subtracted out from the
second symbols of each of the $k$ nodes. This leaves us with the
values of $\underline{g}^t\underline{p}_i$ at $k$ different values
of $i$. Using these, values of $g_0,\ldots,g_{k-1}$ can be
recovered.

Thus all $B$ data units can be recovered by a DC which
connects to any $k$ nodes. We also see that reconstruction is
possible irrespective of the values of the auxiliary vectors
$\underline{u}_i$.

\subsection{Regeneration:}
In our construction, when a node fails, the main vector of the
regenerated node has the same value as that of the failed node,
although the auxiliary vector is allowed to be different. Suppose
node $j$ fails. The node replacing it would contain
$(\underline{f}^t\underline{p}_j \; ,
\;\underline{g}^t\underline{p}_j+\underline{f}^t\underline{\tilde{u}}_j)$
where elements of $\underline{\tilde{u}}_j$ can take any arbitrary
value from $\mathbb{F}_q$ and are not constrained to be equal to
those of $\underline{u}_j$. As the reconstruction property holds
irrespective of the values of $\underline{u}_j$, the regenerated
node along with the existing nodes has all the desired properties.

For regeneration of a failed node, some $d$ nodes give one
(as $\beta=1$) symbol each formed by a linear combination of the
symbols stored in them. Assume that node $\Lambda_{d+1}$ fails and
nodes $\Lambda_1,\ldots,\Lambda_d$ are used to regenerate it, where
the set $\{ \Lambda_1,\ldots,\Lambda_{d+1} \}$ is some subset of
$\{1,\ldots,n\}$, with all elements distinct.

Let $a_i$ and $b_i$ ($i = 1,\ldots,d$) be the coefficients of the
linear combination for the symbol given out by node $\Lambda_i$. Let
$v_i = a_i(\underline{f}^t\underline{p}_{\Lambda_i}) +
b_i(\underline{g}^t\underline{p}_{\Lambda_i} +
\underline{f}^t\underline{u}_{\Lambda_i})$ be this symbol. Let
$\delta_i$ and $\rho_i$ ($i = 1,\ldots,d$) be the coefficients of
the linear combination used to generate the two symbols of the
regenerated node. Thus the regenerated node will be \bea \left(
\sum_{i=1}^{d} \delta_i v_i \; , \; \sum_{i=1}^{d} \rho_i v_i
\right)
 \eea
Choose $b_i=1 \; (i=1,\ldots,d)$. Now choose $\rho_i \;
(i=1,\ldots,d)$ such that \bea \sum_{i=1}^{d} \rho_i b_i
\underline{p}_{\Lambda_i} = \underline{p}_{\Lambda_{d+1}}
\label{eq:rho} \eea and $\delta_i \; (i=1,\ldots,d)$ such that \bea
\sum_{i=1}^{d} \delta_i b_i \underline{p}_{\Lambda_i} =
\underline{0} \label{eq:delta}\eea Equations (\ref{eq:rho}) and
(\ref{eq:delta}) are sets of $k$ linear equations in $d=k+1$
unknowns each. Since $\underline{p}_{\Lambda_i}$'s form a
$k-$dimensional MDS code these can be solved easily in
$\mathbb{F}_q$. This also ensures that we can find a solution to
equation (\ref{eq:delta}) with all $\delta_i$'s non-zero.
\newline
\newline Now, choose $a_i \; (i=1,\ldots,d)$ such that \bea \sum_{i=1}^{d}
\delta_i (a_i \underline{p}_{\Lambda_i}+b_i
\underline{u}_{\Lambda_i}) = \underline{p}_{\Lambda_{d+1}} \eea i.e
\bea \sum_{i=1}^{d} \delta_i a_i \underline{p}_{\Lambda_i} =
\underline{p}_{\Lambda_{d+1}} - \sum_{i=1}^{d} \delta_i b_i
\underline{u}_{\Lambda_i} \label{eq:a} \eea Equation (\ref{eq:a}) is
a set of $k$ linear equations in $d=k+1$ unknowns which can be
easily solved in $\mathbb{F}_q$. Since none of the $\delta_i \;
(i=1,\ldots,d)$ are zero, the particular choice of
$\underline{p}_{\Lambda_i}$'s used guarantees a solution for $a_i \;
(i=1,\ldots,d)$. Hence, regeneration of any node using any $d$ other
nodes is achieved.

\section{Conclusion}\label{sec:conclude}
In this paper, the notion of Exact Regenerating Codes was introduced in which a
failed node is replaced by a new node which is its exact replica. Optimal Exact Regenerating Codes meet the storage-repair
bandwidth tradeoff and have several advantages such as the absence
of communication overhead and a low runtime processing requirement
in comparison with more general regenerating codes. An explicit
construction of exact regenerating codes for the MBR point with
$d=n-1$ was provided, which is well suited for applications such as
mail servers that call for fast recovery upon failure. Subspace
viewpoint was used to prove the uniqueness of our code. At the MSR
point, an explicit construction for regenerating codes for $d=k+1$
was given, that is suitable for peer-to-peer storage systems where
the amount of data stored in each node is to be minimized and where
the number of nodes in the system varies with time. The codes given
for both end points of the storage-repair bandwidth tradeoff have a
low field size requirement and are of low complexity.

\end{document}